\documentstyle[12pt]{article}

\begin{document}

\title{Magnetic field of a neutron star with color superconducting quark 
matter core}
\author{D.M. Sedrakian$^1$, D. Blaschke$^2$ \\
$^1$Physics Department, Yerevan State University\\
Alex Manoogian str. 1, 375025 Yerevan, Armenia\\
$^2$Fachbereich Physik, Universit\"at Rostock\\
Universit\"atsplatz 1, D-18051 Rostock, Germany\\
$^2$Bogoliubov Laboratory for Theoretical Physics, \\
Joint Institute for Nuclear Research, 14 19 80 Dubna, Russia}
\date{}
\maketitle

\begin{abstract}
The behaviour of the magnetic field of a neutron star with a superconducting
quark matter core is investigated in the framework of the Ginzburg-Landau
theory. We take into account the simultaneous coupling of the diquark
condensate field to the usual magnetic and to the gluomagnetic gauge fields.
We solve the Ginzburg-Landau equations by properly taking into account the
boundary conditions, in particular, the gluon confinment condition. We found
the distribution of the magnetic field in both the quark and hadronic phases
of the neutron star and show that the magnetic field penetrates into the
quark core in the form of quark vortices due to the presence of Meissner
currents.
\end{abstract}


\vspace{3cm}

\noindent
MPG-VT-UR 227/02\newline
February 2002 \newpage

\section{ Introduction}

Recently, possible formation of diquark condensates in QCD at finite density
has been re-investigated in series of papers following Refs.\cite{1},\cite{2}.
It has been shown that in chiral quark models with non-perturbative
4-point interaction motivated from instantons \cite{3} or non-perturbative
gluon propagators \cite{4}, \cite{5} the anomalous quark pair amplitudes in
the color antitriplet channel can be very large: of the order of 100 MeV.
Therefore, one expects the diquark condensate to dominate the physics at
densities beyond the deconfinement/chiral restoration transition density and
below the critical temperature (of the order of 50 MeV). Various phases are
possible. The so called two-flavor (2SC) and three-flavor (3SC) phases allow
for unpaired quarks of one color. It has been also found \cite{6}, \cite{7}
that there can exist a color-flavor locked (CFL) phase for not too large
strange quark masses \cite{8}, where color superconductivity is complete in
the sense that diquark condensation results in a pairing gap for the quarks
of all three flavors and colors.

The high-density phases of QCD at low temperatures are relevant for the
explanation of phenomena in rotating massive compact stars which might
manifest themselves as pulsars. Physical properties of these objects (once
being measured) could constrain our hypotheses about the state of matter at
the extremes of densities. In contrast to the situation for the cooling
behaviour of compact stars \cite{9}, \cite{9a}, where the CFL phase is
dramatically different from the 2SC and 3SC phases, we don't expect
qualitative changes of the magnetic field structure for these phases.
Therefore, below we shall restrict ourselves to the discussion of the
simpler two-flavor theory first. Bailin and Love \cite{10} used a
perturbative gluon propagator which yielded a very small pairing gap and
they concluded that quark matter is a type I superconductor, which expells
the magnetic field of a neutron star within time-scales of 10$^4$ years. If
their arguments would hold in general, the observation of life-times for
magnetic fields as large as 10$^7$years \cite{11}, \cite{12} would exclude
the occurence of an extended superconducting quark matter core in pulsars.
These estimates are not valid for the case of diquark condensates
characterized by large quark gaps. Besides, in Ref. \cite{13} the authors
found that within recent non-perturbative approaches for the effective quark
interaction that allow for large pairing gaps the quark condensate forms a
type II superconductor. Consequently for the magnetic field $H<H_{c1}$ there
exists a Meissner effect and for $H_{c2}>H>H_{c1}$ the magnetic field can
penetrate into the quark core in quantized flux tubes. However, they have
not considered in that paper the simultaneous coupling of the quark fields
to the magnetic and gluomagnetic gauge fields.

Though color and ordinary electromagnetism are broken in a color
superconductor, there is a linear combination of the photon and the gluon
that remains massless. The authors of Ref. \cite{14} have considered the
problem of the presence of magnetic fields inside color superconducting
quark matter taking into account the possibility of the so called ''rotated
electromagnetism''. They came to the conclusion that there is no Meissner
effect and the external static homogeneous magnetic field can penetrate into
superconducting quark matter because in their case it obeys the sourceless
Maxwell equations. To our opinion this result is obtained when one does not
pose correct boundary conditions for the fields. Obviously it is
energetically favorable to expell the magnetic field rather than to allow
its penetration inside the superconducting matter. Using for the description
of the diquark condensate interacting with two gauge fields the same model
as in Refs. \cite{8,9,14,15}, the authors of Ref. \cite{15a} have shown 
that the presence of the massless excitation in 
the spectrum does not prevent the Meissner
currents to effectively screen static, homogeneous, external magnetic fields.
In the present paper we will extend those studies to the consideration of 
inhomogeneous, vortex-type external fields. 

In Ref. \cite{15} we have derived the Ginzburg-Landau equations of
motion for the diquark condensate placed in static magnetic and gluomagnetic
fields, 
\begin{equation}
ad_p+\beta (d_pd_p^{*})d_p+\gamma 
(i\bigtriangledown -\frac e3 \stackrel{\rightarrow }{A}
+\frac g{\sqrt{3}}\stackrel{\rightarrow }{G}_8)^2d_p =0,
\label{ku}
\end{equation}
where $d_p$ is the order parameter, $a=t$ $dn/dE,\beta =(dn/dE)$ $7$ $\zeta
(3)(\pi T_c)^{-2}/8,$ $\gamma =p_F^2$ $\beta /(6\mu ^2),$ $dn/dE=p_F$ $\mu
/\pi ^2$, $t=(T-T_c)/T_c$, $T_c$ being the critical temperature, $p_F$-the
quark Fermi momentum, and for the gauge fields 
\begin{equation}
\lambda _q^2{\rm rot}~{\rm rot}\stackrel{\rightarrow }{A}+\sin ^2\alpha 
\stackrel{\rightarrow }{A}=i\frac{\sin \alpha (d_p\nabla
d_p^{*}-d_p^{*}\nabla d_p)}{2qd^2}+\sin \alpha \cos \alpha 
\stackrel{\rightarrow }{G}_{8,}  
\label{ka}
\end{equation}
\begin{equation}
\lambda _q^2{\rm rot}~{\rm rot}\stackrel{\rightarrow }{G}_8+\cos ^2\alpha 
\stackrel{\rightarrow }{G}_8=-i\frac{\cos \alpha (d_p\nabla
d_p^{*}-d_p^{*}\nabla d_p)}{2qd^2}+\sin \alpha \cos \alpha 
\stackrel{\rightarrow }{A}.  \label{kq}
\end{equation}
These equations introduce a ''new'' charge of the diquark pair 
$q=\sqrt{\eta^2e^2+g^2}P_8$, $P_8=1/\sqrt{3}$, 
and for the diquark condensate with
paired blue-green and green-blue $ud$ quarks one has $\eta =1/\sqrt{3}$. 
The penetration depth of the magnetic and gluomagnetic fields $\lambda _q$ 
and the mixing angle $\alpha $ are given by 
\begin{equation}
\lambda _q^{-1}=qd \sqrt{2\gamma },\;\,\cos \alpha 
=\frac g{\sqrt{\eta^2e^2+g^2}}~.  
\label{kb}
\end{equation}
At neutron star densities gluons are strongly coupled ($g^2/4\pi \sim 1)$
whereas photons are weakly coupled $(e^2/4\pi =1/137)$, so that $\alpha
\simeq \eta e/g$ is small. For $g^2/4\pi \simeq 1$ we get $\alpha \simeq
1/20 $. The new charge $q$ is by an order of magnitude larger than 
$e/\sqrt{3}$.

Please notice also that since red quarks are normal in the 2SC and 3SC
phases, there exist the corresponding normal currents $j_\mu ^r(A)=-\Pi
_{\mu \nu }^{el}A^\nu $ and $j_\mu ^r(G_8)=-\Pi _{\mu \nu }^{gl}G_8^\nu $
which however do not contribute in the static limit under consideration to
the above Ginzburg - Landau equations, cf. \cite{BS93}. Thus, the
qualitative behavior of the static magnetic field for all three 2SC, 3SC,
and CFL phases is the same. Recently the influence of a constant uniform
chromomagnetic field on the formation of color superconductivity and the
role of the Meissner effect for tightly bound states have been considered in
the papers \cite{16}, \cite{17}.

We will solve the Ginzburg-Landau equations (\ref{ku}), (\ref{ka}), 
(\ref{kq}) by properly taking into account the boundary conditions, in 
particular, the gluon confinment condition. 
We will find the distribution of the
magnetic field in both the quark and hadronic phases of the neutron star and
will show that the magnetic field penetrates into the quark core in the form
of quark vortices due to the presence of Meissner currents.

We assume a sharp boundary between the quark and hadron matter since the
diffusion boundary layer is thin, of the order of the size of the
confinement radius ($\sim 0.2\div 0.4$ fm), and we suppose that the
coherence length $l_{\xi}=\sqrt{\gamma /(-2a)}$ is not less than this value
and the magnetic and gluomagnetic field penetration depth $\lambda_q$ is
somewhat larger than the confinement radius. Also we assume that the size of
the quark region is much larger than all mentioned lengths.

\section{ Solution of Ginzburg-Landau equations}

\bigskip Let us rewrite equations (\ref{ka}) and (\ref{kq}) for a
homogeneous superconducting matter region being a type II superconductor in
the following form 
\begin{eqnarray}
\lambda _q^{2\;}{\rm rot\;rot}\vec{A}+\sin ^2\alpha \vec{A} &=&\vec{f}~\sin
\alpha +\sin \alpha \cos \alpha \vec{G}_8,  \label{kc} \\
\lambda _q^2~{\rm rot}~{\rm rot}\vec{G}_8+\cos ^2\alpha \vec{G}_8 &=&-\vec{f}
~\cos \alpha +\sin \alpha \cos \alpha \vec{A}~,  \label{kd}
\end{eqnarray}
where 
\begin{equation}
\vec{f}=\lambda _q^2~4\pi iq\gamma \left[ \vec{d}\vec{\nabla}\vec{d}^{*}-
\vec{d}^{*}\vec{\nabla}\vec{d}\right]
\end{equation}
and obeys the equation 
\begin{equation}
{\rm rot}{\rm rot}\vec{f}=0~.  \label{f}
\end{equation}
If we introduce 
\begin{equation}
\vec{A}^{\prime }=\vec{A}-\frac{\vec{f}}{2\sin \alpha }~,~~~~
\vec{G}_8^{\prime }=\vec{G}_8+\frac{\vec{f}}{2\cos \alpha }~,  
\label{prime}
\end{equation}
then Eqs. (\ref{kc}) and (\ref{kd}) can be rewritten in the form: 
\begin{eqnarray}
\frac{\lambda _q^{2\;}}{\sin \alpha }{\rm rot\;rot}\vec{A}^{\prime } 
&=&\cos \alpha \vec{G}_8^{\prime }-\sin \alpha \vec{A}^{\prime }~,  
\label{kc'} \\
-\frac{\lambda _q^2}{\cos \alpha }~{\rm rot}~{\rm rot}\vec{G}_8^{\prime }
&=&\cos \alpha \vec{G}_8^{\prime }-\sin \alpha \vec{A}^{\prime }~,
\label{kd'}
\end{eqnarray}
We can define $\vec{G}_8^{\prime }$ from (\ref{kc'}) as follows 
\begin{equation}
\vec{G}_8^{\prime }=\frac{\lambda _q^2~{\rm rot}~{\rm rot}\vec{A}^{\prime
}+\sin ^2\alpha \vec{A}^{\prime }}{\sin \alpha \cos \alpha }.  \label{ke}
\end{equation}
We derive from equations (\ref{kd'}) and (\ref{ke}) the relation 
\begin{equation}
{\rm rot}~{\rm rot}\vec{G}_8^{\prime }=-\cot \alpha ~{\rm rot}~{\rm rot}
\vec{A}^{\prime }~.  \label{kf}
\end{equation}
We can consider instead of the system of coupled equations (\ref{kc'}) and 
(\ref{kd'}) the equivalent system (\ref{ke}) and (\ref{kf}). 
We substitute 
$\vec{G}_8^{\prime }$ from (\ref{ke}) into (\ref{kf}) and obtain the
following equation 
\begin{equation}
{\rm rot}~{\rm rot}~(\lambda _q^2~{\rm rot}~{\rm rot}\vec{A}^{\prime} +
\vec{A}^{\prime })=0.  \label{kg}
\end{equation}
We introduce the new function $\vec{M}^{\prime }$ as 
\begin{equation}
\vec{M}^{\prime }={\rm rot~rot}\vec{A}^{\prime }~,  \label{kh}
\end{equation}
we obtain 
\begin{equation}
\lambda _q^{2~}{\rm rot~rot}\vec{M}^{\prime }+\vec{M}^{\prime }=0.
\label{kl}
\end{equation}
So we can determine the function $\vec{A}^{\prime }$ by simultaneous
solution of the equations (\ref{kh}) and (\ref{kl}). Then we can find the
electromagnetic potential $\vec{A}$ and the gluonic potential $\vec{G}_8$
from equations (\ref{ke}) and (\ref{prime}).

In order to determine the distribution of electromagnetic and gluonic
potentials inside the superconducting quark matter core in response to an
external magnetic field we shall require on the quark-hadronic matter
boundary both the continuity of the magnetic field and the vanishing of the
gluon potential ($\vec{G}_8=0)$ due to gluon confinement. As we shall see
below, these conditions are sufficient for a unique determination of the
magnetic and gluomagnetic fields inside the quark matter region.

We shall assume that a neutron star with radius $R$ possesses a spherical
core of radius $a$ consisting of color superconducting quark matter
surrounded by a spherical shell of hadronic matter with thickness $R-a$. 
The functions $\vec{M}^{\prime }$, $\vec{A}$ and $\vec{G}_8$ in spherical
coordinates $(r,\vartheta ,\varphi )$ have only $\varphi $-components 
${M}_\varphi ^{\prime }(r,\vartheta )$, $\vec{A}_\varphi (r,\vartheta )$ and 
$\vec{G}_{8~\varphi }(r,\vartheta )$. For the solution of the equation (\ref
{kl}) we make the ansatz ${M}_\varphi ^{\prime }(r,\vartheta )={M}_\varphi
(r)~\sin \vartheta $. Then equation (\ref{kl}) can be written as 
\begin{equation}
\frac{d^2M_\varphi (r)}{dr^2}+\frac 2r\frac{dM_\varphi (r)}{dr}-(\frac
2{r^2}+\frac 1{\lambda _q^2})M_\varphi (r)=0~.  \label{c}
\end{equation}
The solution of equation (\ref{c}) is 
\begin{equation}
M_\varphi (r)=\frac 1{r^2}\left[ c_1^{\prime }\left( 1-\frac r{\lambda
_q}\right) {\rm e}^{\frac r{\lambda _q}}+c_2^{\prime }\left( 1+\frac
r{\lambda _q}\right) {\rm e}^{-\frac r{\lambda _q}}\right] ~.  \label{c1}
\end{equation}
The condition that $M_\varphi (r)$ tends to zero at the center of the quark
core gives $c_1^{\prime }=-c_2^{\prime }$ , so that 
\begin{equation}
M_\varphi (r)=\frac{c_1}{r^2}\left[ \sinh \frac r{\lambda _q}-\frac
r{\lambda _q}\cosh \frac r{\lambda _q}\right] ~.  \label{c2}
\end{equation}
Substituting the solution (\ref{c2}) into equation (\ref{kh}) for 
$\vec{A}^{\prime }$ we obtain the following solution 
\begin{equation}
A_\varphi ^{\prime }(r,\vartheta )=M_\varphi ^{\prime }(r,\vartheta
)+c_0^{\prime }~r~\sin \vartheta ~.  \label{e}
\end{equation}
Taking into account (\ref{e}) and using equation (\ref{prime}) for the
electromagnetic potential we obtain 
\begin{equation}
A_\varphi (r,\vartheta )=M_\varphi ^{\prime }(r,\vartheta )+c_0^{\prime
}~r~\sin \vartheta +\frac{f_\varphi (r,\vartheta )}{2~\sin \alpha }~.
\label{e1}
\end{equation}
The unknown function $f_\varphi (r,\vartheta )$ we will find from the
solution of equation (\ref{f}) which gives 
\begin{equation}
f_\varphi (r,\vartheta )=c_0~r~\sin \vartheta ~.
\end{equation}
Finally we get for the electromagnetic potential 
\begin{equation}
A_\varphi (r,\vartheta )=\left[ M_\varphi (r)+c_0^{\prime }r+\frac{c_0r}
{2~\sin \alpha }\right] \sin \vartheta ~.  \label{e2}
\end{equation}
Using equations (\ref{ke}) and (\ref{prime}), we find the gluonic potential 
$G_{8\varphi }$ in the form 
\begin{equation}
G_{8\varphi }(r,\vartheta )=\left[ -\cot \alpha ~M_\varphi (r)+\tan \alpha
~c_0^{\prime }~r-\frac{c_0~r}{2~\sin \alpha }\right] ~\sin \vartheta ~.
\label{g8}
\end{equation}
The constant $c_0^{\prime }$ we can define using the gluon confinement
condition on the surface of the quark matter core $G_{8\varphi }(a,\vartheta
)=0$~. This condition will define $c_0^{\prime }$ as 
\begin{equation}
c_0^{\prime }=\cot ^2\alpha \frac{M_\varphi (a)}a+\frac{c_0}{2~\sin \alpha }
~,
\end{equation}
to be substituted into equations (\ref{e2}) and (\ref{g8}) so that we obtain
for the final expressions for the electromagnetic and gluonic potentials 
\begin{eqnarray}
A_\varphi (r,\vartheta ) &=&\left[ M_\varphi (r)+\cot ^2\alpha \frac
raM_\varphi (a)+\frac{c_0~r}{\sin \alpha }\right] ~\sin \vartheta ~,
\label{m1} \\
G_{8\varphi }(r) &=&-\left[ M_\varphi (r)-\frac raM_\varphi (a)\right] \cot
\alpha ~\sin \vartheta ~.  \label{m2}
\end{eqnarray}
In ending this section we mention that the electromagnetic potential in the
hadronic matter part of the neutron star can be found from the solution (\ref
{c1}) by replacing the penetration depth for quark matter $\lambda _q$ with
that for hadronic matter $\lambda _p$~.

\section{The magnetic field components for the neutron star}

The components of the magnetic fields in quark and hadronic matter can be
found from those of the vector potentials using the formula $\vec{B} = 
{\rm rot}~\vec{A}$~. In spherical coordinates we have 
\begin{eqnarray}  \label{ba}
B_r &=& \frac{1}{r~\sin \vartheta} \frac{\partial}{\partial \vartheta}
\left(\sin \vartheta~ A_\phi (r,\vartheta) \right)~, \\
B_\vartheta &=& - \frac{1}{r} \frac{\partial}{\partial r} \left(r~ A_\phi
(r,\vartheta) \right)~.  \label{bb}
\end{eqnarray}
For the case of quark matter we have to insert into these formulae the
expression (ref{m1}) for $A_\phi (r,\vartheta)$. Then finally we get (for 
$r\le a$) 
\begin{eqnarray}  \label{bqa}
B_r^q &=& \left[ \frac{2 M_\varphi (r)}{r} +2 \cot^2 \alpha \frac{M_\varphi
(a)}{a} + \frac{2 c_0}{\sin \alpha}\right]~\cos \vartheta~, \\
B_\vartheta^q &=& - \left[\frac{1}{r} \frac{d}{d r} \left(r~M_\varphi
(r)\right) +2 \cot^2 \alpha \frac{M_\varphi (a)}{a} + \frac{2 c_0}{\sin
\alpha} \right] \sin \vartheta ~,  \label{bqb}
\end{eqnarray}
where $M_\varphi (r)$ is defined by equation (\ref{c2}).

The magnetic field in hadronic matter phase we can find from the solution 
(\ref{c1}) by taking into account that proton vortices in this phase generate
a homogeneous mean magnetic field with amplitude $B$ and direction parallel
to the axis of rotation of the star \cite{15a}. For the components of the
magnetic field $\vec{B}^p$ in the hadronic phase (for $a\le r\le R$) we get
the following expressions 
\begin{eqnarray}
B_r^p &=&\left[ \frac{2A_\phi (r)}r+B\right] \cos \vartheta ~,  \label{bpa}
\\
B_\vartheta ^p &=&-\left[ \frac 1r\frac d{dr}\left( r~A_\phi (r)\right)
+B\right] \sin \vartheta ~,  \label{bpb}
\end{eqnarray}
where 
\begin{equation}
A_\varphi (r)=\frac{c_2}{r^2}\left( 1-\frac r{\lambda _p}\right) {\rm e}
^{\frac r{\lambda _p}}+\frac{c_3}{r^2}\left( 1+\frac r{\lambda _p}\right) 
{\rm e}^{-\frac r{\lambda _p}}~.  \label{c3}
\end{equation}
As we have mentioned above $\lambda _p$ is the penetration depth in hadronic
matter.

The external magnetic field $\vec{B}^e$ in the region outside of the star 
($r\ge R$) has to be dipolar and their components are given by the following
expressions 
\begin{equation}
B_r^e=\frac{2{\cal M}}{r^3}\cos \vartheta ~,~~~B_\vartheta ^e=\frac{{\cal M}
}{r^3}\sin \vartheta ~,  \label{be}
\end{equation}
where ${\cal M}$ is the full magnetic moment of the star. The unknown
constants $c_0$, $c_1$, $c_2$, $c_3$ and ${\cal M}$ in equations (\ref{ba})-(
\ref{be}) have to be defined from the continuity conditions of the magnetic
field components at $r=a$ and $r=R$ and from the condition 
\begin{equation}
B^q~V_1+B~V_2={8\pi /}{3}{\cal M}~,  \label{32}
\end{equation}
where $B^q$ is the z component of the magnetic field in the quark matter
region with the volume $V_1$. $V_2$ is the volume of the hadronic matter
region. Here we suppose that the magnetic field in both regions is mainly
constant and parallel to the axis of rotation $z$. As we will see later this
supposition is fulfilled with high accuracy because $a$, $R$ and $R-a$ are
much greater than $\lambda _p$ and $\lambda _q$.

The continuity of the magnetic field components at $r=a$ and $r=R$ gives us
the following equations 
\begin{eqnarray}
\frac{2M_\varphi (a)}a+2\cot ^2\frac{M_\varphi (a)}a+\frac{2c_0}{\sin \alpha 
} &=&\frac{2A_\phi (a)}a+B~,  \label{33c} \\
\frac 1r\frac d{dr}\left( r~M_\phi (r)\right) \bigg| _{r=a}+2\cot ^2\frac{
M_\varphi (a)}a+\frac{2c_0}{\sin \alpha } &=&\frac 1r\frac d{dr}\left(
r~A_\phi (r)\right) \bigg| _{r=a}+B~, \\
\frac{2A_\phi (R)}R+B &=&\frac{2{\cal M}}{R^3}~, \\
\frac 1r\frac d{dr}\left( r~A_\phi (r)\right) \bigg| _{r=R}+B &=&-\frac{
{\cal M}}{R^3}~.  \label{33d}
\end{eqnarray}
Substituting the functions $M_\varphi (r)$ and $A_\varphi (r)$ from
equations (\ref{c2}) and (\ref{e1}) into the system of equations (\ref{33c})
- (\ref{33d}) we find the following system of equations 
\begin{eqnarray}
0 &=&c_1\left[ \left( 1+\frac{a^2}{3\lambda _q^2}\right) \sinh \frac
a{\lambda _q}-\frac a{\lambda _q}\cosh \frac a{\lambda _q}\right] -c_2\left[
1-\frac a{\lambda _p}+\frac{a^2}{3\lambda _p^2}\right] {\rm e}^{\frac
a{\lambda _p}}  \nonumber  \label{34c} \\
&&-c_3\left( 1+\frac a{\lambda _p}+\frac{a^2}{3\lambda _p^2}\right) {\rm e}
^{-\frac a{\lambda _p}}~, \\
D &=&c_1\left( \sinh \frac a{\lambda _q}-\frac a{\lambda _q}\cosh \frac
a{\lambda _q}\right)   \nonumber \\
&&-\left[ c_2\left( 1-\frac a{\lambda _p}\right) {\rm e}^{\frac a{\lambda
_p}}+c_3\left( 1+\frac a{\lambda _p}\right) {\rm e}^{-\frac a{\lambda
_p}}\right] \sin ^2\alpha ~, \\
{\cal M} &=&c_2\left( 1-\frac R{\lambda _p}+\frac{R^2}{3\lambda _p^2}\right) 
{\rm e}^{\frac R{\lambda _p}}+c_3\left( 1+\frac R{\lambda _p}+\frac{R^2}{
3\lambda _p^2}\right) {\rm e}^{-\frac R{\lambda _p}}~, \\
{\cal M} &=&c_2\left( 1-\frac R{\lambda _p}\right) {\rm e}^{\frac R{\lambda
_p}}+c_3\left( 1+\frac R{\lambda _p}\right) {\rm e}^{-\frac R{\lambda _p}}+
\frac{B~R^3}2~,  \label{34d}
\end{eqnarray}
where 
\begin{equation}
D=\frac{Ba^3}2\sin ^2\alpha -c_0a^3\sin \alpha ~.  \label{36}
\end{equation}
The solution of this system of equations using the fact that $a$, $R$ and 
$R-a$ are much larger than $\lambda _q$ and $\lambda _p$ we obtain the
following expressions for $c_1$, $c_2$, $c_3$ and ${\cal M}$ 
\begin{eqnarray}
c_1 &=&-\frac{\lambda _q^2}{a\lambda _p}~~\frac D{\sin ^2\alpha +\frac{
\lambda _q}{\lambda _p}}~~\frac 1{\sinh \frac a{\lambda _q}}~  \label{35c} \\
c_2 &=&\frac{\lambda _p}{2a}~~\frac{D~{\rm e}^{-\frac R{\lambda _p}}}{\sin
^2\alpha +\frac{\lambda _q}{\lambda _p}}~~\frac 1{\sinh \frac{R-a}{\lambda_p
}}~ \\
c_3 &=&-\frac{\lambda _p}{2a}~~\frac{D~{\rm e}^{\frac R{\lambda _p}}}{\sin
^2\alpha +\frac{\lambda _q}{\lambda _p}}~~\frac 1{\sinh \frac{R-a}{\lambda_p
}}~ \\
{\cal M} &=&\frac{B~R^3}2-\frac Ra~~\frac D{\sin ^2\alpha +\frac{\lambda _q}{
\lambda _p}}~~\frac 1{\sinh \frac{R-a}{\lambda _p}}~.  \label{35d}
\end{eqnarray}
In order to obtain a final result we have to determine the constant $c_0$
assuming $B$ is constant. The latter is given since in the hadronic phase
entrainment currents generate proton vortices which provide this field \cite
{DSed}.

Before we go over to the determination of $c_0$ let us consider the
behaviour of the magnetic field at distances $r$ much larger than $\lambda _p
$ and $\lambda _q$. Also we take into account that $\lambda \ll \lambda
_p\ll a$, $\lambda \ll \lambda _p\ll R$ and $\lambda \ll \lambda _p\ll R-a$.
Using the expressions (\ref{35c}) - (\ref{35d}) for the constants, which we
have obtained within this approximation, the components of the magnetic
field in the different regions of the neutron star take the following form. 
\newline
For $r\le a$ 
\begin{eqnarray}
B_r^q &=&\left[ \frac{2D}{ar^2}\frac{{\rm e}^{-\frac{a-r}{\lambda _q}}}{\sin
^2\alpha +\frac{\lambda _q}{\lambda _p}}\frac{\lambda _q}{\lambda _p}+2\cot
^2\alpha \frac{M_\varphi (a)}a+\frac{2c_0}{\sin \alpha }\right] \cos
\vartheta ~  
\label{37a} \\
B_\vartheta ^q &=&-\left[ \frac D{ar\lambda _p}\frac{{\rm e}^{-\frac{a-r}{
\lambda _q}}}{\sin ^2\alpha +\frac{\lambda _q}{\lambda _p}}+2\cot ^2\alpha 
\frac{M_\varphi (a)}a+\frac{c_0}{\sin \alpha }\right] \sin \vartheta ~,
\label{37b}
\end{eqnarray}
for $a\le r\le R$ 
\begin{eqnarray}
B_r^p &=&\left[ -\frac{2D}{ar^2}\frac{{\rm e}^{-\frac{r-a}{\lambda _p}}}{
\sin ^2\alpha +\frac{\lambda _q}{\lambda _p}}\frac{\lambda _q}{\lambda _p}
+B\right] \cos \vartheta ~,  
\label{37c} \\
B_\vartheta ^p &=&-\left[ \frac D{ar\lambda _p}\frac{{\rm e}^{-\frac{r-a}{
\lambda _p}}}{\sin ^2\alpha +\frac{\lambda _q}{\lambda _p}}+B\right] \sin
\vartheta ~,  
\label{37d}
\end{eqnarray}
and for $r\ge R$ 
\begin{equation}
B_r^e=\frac{B~R^3}{r^3}\cos \vartheta ~,~~~B_\vartheta ^e
=\frac{B~R^3}{2r^3} \sin \vartheta ~,  
\label{37e}
\end{equation}
As can be seen from the obtained solutions, the magnetic field in both quark
and hadronic phases depends on $r$ only close to the phase boundary at $r=a$
. In more detail, the part of the solution for the field which depends on $r$
tends to zero in the quark core at distances $a-r\gg \lambda _q$ and in the
hadronic shell at distances $r-a\gg \lambda _p$. But since $\lambda _p$ and 
$\lambda _q$ are much smaller than $a$ and $R-a$ except a small layer at the
surface of the quark core with the depth of the order 
$\lambda _p+\lambda _q$, the magnetic field in both phases is constant and 
directed parallel to the
rotation axis of the star, see solution (\ref{37a}) - (\ref{37d}). 
We conclude that in the main part of the volume of the quark and hadron phases
the magnetic field is constant and has the direction $z$. 
In this approximation the condition (\ref{32}) is satisfied. 
Therefore inserting in 
(\ref{32}) the relation ${\cal M}=B~R^3/2$, see equation (\ref{37e}), we have 
\begin{equation}
B^q=2\cot ^2\alpha \frac{M_\varphi (a)}a+\frac{2~c_0}{\sin \alpha }=B~.
\label{38}
\end{equation}
Solving equations (\ref{36}) and (\ref{38}) we finally obtain 
\begin{equation}
c_0=\frac{B~\sin \alpha }2~,~~~~D=0~.  \label{39}
\end{equation}
Thus in this approximation the magnetic field $\vec{B}$ enters from the
hadronic phase into the quark phase in the form of quark vortices \cite{15}.
The transition zone is of the order $\lambda _p+\lambda _q$ which entails
that the quantity $D$ is small of the order $(\lambda _p+\lambda _q)/a$ so 
that the condition $D=0$ is well fulfilled.

\section{Conclusion}

We have investigated the behaviour of the magnetic field of a neutron star
with superconducting quark matter core in the framework of the
Ginzburg-Landau theory whereby the simultaneous coupling of the diquark
condensate field to the usual magnetic and to the gluomagnetic gauge fields
has been taken into account (rotated electromagnetism). In solving the
Ginzburg-Landau equations for this problem, we have respected the boundary
conditions properly, in particular, the gluon confinement condition. We have
found the distribution of the magnetic field in both the quark and hadronic
phases of the neutron star and have shown that the magnetic field penetrates
into the quark core in the form of quark vortices due to the presence of
Meissner currents.

\subsection*{Note added:}

After acceptance of this paper for publication in {\it Astrofizika}, the 
paper \cite{iida} has appeared on the e-print Archive, which complements the 
present study and adds a detailed discussion of the effects of rotation on
vortices and gluon-photon mixing.  

\subsection*{Acknowledgement}

D.B. and D.S. acknowledge the support from DAAD for mutual visits and the
hospitality extended to them at the partner Universities in Yerevan and
Rostock, respectively. D.S. acknowledges FAR/ANSEF support by the Yerevan
University ANSEF grant {\it No} PS 51-01.

\end{document}